\documentclass[useAMS,usenatbib]{mn2e}
\usepackage{amssymb}
\usepackage{epsfig}
\usepackage{color}

\title[Modeling the light curves PSR B1259-63/SS 2883]{Modeling the multi-wavelength light curves of PSR
B1259-63/SS 2883}
\author[S. W. Kong, Y. W. Yu, Y. F. Huang and K. S. Cheng]{S. W. Kong$^{1}$, Y. W.
Yu$^{2}$, Y. F. Huang$^{1,3}$ and K. S. Cheng$^{4}$\thanks{E-mail: hrspksc@hkucc.hku.hk (KSC)}\\
$^{1}$Department of Astronomy, Nanjing University,
Nanjing 210093, China\\
$^{2}$Institute of Astrophysics, Huazhong
Normal University, Wuhan 430079, China\\
$^{3}$Key Laboratory of
Modern Astronomy and Astrophysics (Nanjing University), Ministry of
Education, China\\
$^{4}$Department of Physics, The University of Hong Kong, Pokfulam
Road, Hong Kong, China}

\begin{document}

\date{Accepted ? December ?. Received ? December ?; in original form ? October ?}

\pagerange{\pageref{firstpage}--\pageref{lastpage}} \pubyear{2011}

\maketitle

\label{firstpage}

\begin{abstract}
PSR B1259-63/SS 2883 is a binary system in which a 48-ms pulsar
orbits around a Be star in a high eccentric orbit with a long
orbital period of about 3.4 yr. Extensive broadband observational
data are available for this system from radio band to very high
energy (VHE) range. The multi-frequency emission is unpulsed and
nonthermal, and is generally thought to be related to the
relativistic electrons accelerated from the interaction between the
pulsar wind and the stellar wind, where X-ray emission is from the
synchrotron process and the VHE emission is from the inverse Compton
(IC) scattering process. Here a shocked wind model with variation of
the magnetization parameter $\sigma$ is developed for explaining the
observations. By choosing proper parameters, our model could
reproduce two-peak profile in X-ray and TeV light curves. The effect
of the disk exhibits an emission and an absorption components in the
X-ray and TeV bands respectively. We suggest that some GeV flares
will be produced by Doppler boosting the synchrotron spectrum. This
model can possibly be used and be checked in other similar systems
such as LS I+61\textordmasculine 303 and LS 5039.
\end{abstract}


\section{Introduction}

The discovery of the PSR B1259-63/SS 2883 system was first reported
in 1992 (Johnston et al. 1992), and it is a binary system containing
a rapidly rotating pulsar PSR B1259-63 in orbit around a massive Be
star companion SS 2883. The spin period of the pulsar is $P$ = 47.76
ms and the spindown luminosity is $L_{\rm sp} \simeq 8 \times
10^{35}$ ${\rm ergs}$ ${\rm s}^{-1}$. The distance between the
system and Earth is estimated to be about 1.5 kpc (Johnston et al.
1994).

This system is special for having unpulsed and nonthermal multi-band
observational data spanning from radio band to TeV range, which are
variable along with the orbital phase (Johnston et al. 1994, 1996,
2005; Chernyakova et al. 2006, 2009; Uchiyama et al. 2009; Aharonian
et al. 2005, 2009). The broadband emission from this system is
generally believed to be produced by the ultrarelativistic electrons
accelerated in the shock from the interaction between the pulsar
wind and the outflow of Be star (Tavani \& Arons 1997; Kirk, Ball \&
Skj{\ae}raasen 1999; Khangulyan et al. 2007; Takata \& Taam 2009;
Kerschhaggl 2010). Massive stars usually produce very strong stellar
wind during their life time. For a Be star, the stellar wind is
composed of two parts: a polar component and an equatorial
component. The polar wind is faster and thinner; the equatorial
component is slower and denser, and forms a disk surrounding the Be
star. The interaction between the pulsar wind and the stellar
outflow will form a termination shock at the position where the
dynamical pressures of the pulsar wind and the stellar wind are in
balance, and this shock will accelerate electrons to relativistic
velocities. These accelerated electrons will emit nonthermal
emission via synchrotron process for X-ray regime (Tavani \& Arons
1997) or IC [either synchrotron self-Compton (SSC) or external
inverse Compton scattering the thermal photons from the Be star
(EIC)] process for VHE regime (Kirk, Ball \& Skj{\ae}raasen 1999).
So this system is an important astrophysical laboratory for studying
the interaction between the pulsar wind and the stellar wind, and
the nonthermal radiation arising from the shocked pulsar wind.

However, the prediction from a simple wind interaction model is not
consistent with the observations. The simple wind interaction model
predicts that the X-ray light curve reaches a maximum in flux at
periastron (Tavani \& Arons 1997), but the observed X-ray light
curve has a two-peak profile and has a drop in photon flux around
the periastron (Kaspi et al. 1995; Hirayama et al. 1999; Chernyakova
et al. 2006, 2009). The behavior of TeV light curve is similar to
that in X-ray range (Aharonian et al. 2005, 2009). Some authors
suggest that hadronic scenarios may take place in the system
(Kawachi et al. 2004; Neronov \& Chernyakova 2007). In the
hadronic model, The Ultra-high energy (UHE) protons will be produced
by the interaction between the pulsar wind and the dense equatorial
wind. These UHE protons can loose their energy in interactions with
the protons from the stellar wind and produce pions. The decay of
neutral pions will produce the TeV emission, and the leptons from
the decay of charged pions will produce the radio to GeV emission
via the synchrotron and IC processes respectively. In this
situation, the two-peak profile in the light curve is related to the
passage of the Be star disk. On the other hand, some other authors
use some revised leptonic models to explain the drop of photon flux
towards periastron (Kangulyan et al. 2007; Takata \& Taam 2009).
Kangulyan et al. (2007) suggest that the observed TeV light curve
can be explained (i) by increasing of the adiabatic loss rate close
to periastron or (ii) by the early sub-TeV cut-offs in the
energy spectra of electrons due to the enhanced rate of Compton
losses near the periastron. More recently, Takata \& Taam (2009)
explain the observed light curves by varying the microphysical
parameters in two different ways: (i) they use a given power-law
spectral index $p$ of electron distribution, and vary the
magnetization parameter $\sigma$ and the pulsar wind bulk Lorentz
factor $\Gamma$; (ii) the magnetization parameter and the power law
index are varied for a given bulk Lorentz factor.

Among all the mechanisms, we believe that the variation of
microphysics model needs to be paid particular attention, especially
the variation of the magnetization parameter $\sigma$. The varying
of the magnetization parameter $\sigma$ along with the distance from
the pulsar is a very natural hypothesis. In the Crab Nebula, $\sigma
\sim 0.003$ at a distance of $r_{\rm s} \sim 3 \times 10^{17}$ cm
(Kennel \& Coroniti 1984a, 1984b). But at the light cylinder of a
pulsar, $\sigma$ should be much larger and is about $10^4$ - $10^5$.
Between the light cylinder and the termination shock, the magnetic
energy will be gradually converted into the particle kinetic energy,
and $\sigma$ will vary along with the energy transition. Especially,
a larger $\sigma$ around periastron will decrease the energy in
particles and make the cooling faster, which may cause a drop in
photon flux.

The PSR B1259-63/SS 2883 system is also special for having a large
eccentricity of $e = 0.87$ and a long orbital period of about 3.4
yr. The semimajor axis of the orbit is about 5 AU. Although the wind
interaction model for this binary system is very similar to the
pulsar wind nebulae (PWN) model (Kennel \& Coroniti 1984a, 1984b),
the detailed situation is different. For the PWN around an isolated
pulsar like Crab, the distance between the pulsar and the
termination shock $r_{\rm s}$ is about 0.1 pc and the magnetic
parameter $\sigma$ is about 0.003 (Kennel \& Coroniti 1984a, 1984b).
However, in the case of PSR B1259-63/SS 2883 system, due to the
highly eccentric orbit, $r_{\rm s}$ is in the range of 0.1 - 1 AU
and $\sigma$ should be accordingly larger. This gives us the unique
opportunity to study the physical properties of the pulsar wind near
the light cylinder where $\sigma$ is much higher.

In this paper we develop a simple shocked wind model, in which the
magnetization parameter $\sigma$ varies with $r_{\rm s}$ to
reproduce the broadband observational data. Besides PSR B1259-63/SS
2883, there are also some other similar binary systems being found
for emitting the X-ray and VHE emission, such as LS
I+61\textordmasculine 303 (Albert et al. 2006) and LS 5039
(Aharonian et al. 2006). We suggest that this model may also be used
in these similar systems. The outline of our paper is as follows: in
Section 2, we introduce our model in detail. We then present our
results and the comparison with observations in Section 3. Our
discussion and conclusion are presented in Section 4.

\section{Model description}

In our model, the broadband emission of PSR B1259-63/SS 2883 system
is mainly from the shock-accelerated electrons. Due to the
interaction between the pulsar wind and the stellar wind, strong
shocks will be formed, and the electrons will be accelerated at the
shock front of the pulsar wind. This shock will also compress the
magnetic field in the pulsar wind. These shocked relativistic
electrons move in the magnetic field and the photon fields (either
from the synchrotron emission or from the Be star), and emit
synchrotron and IC radiation to produce the multi-band emission.

\subsection{Stellar and pulsar winds, and termination shock}

The companion star of PSR B1259-63, which is named SS 2883, is a
massive Be type star. Be stars are characterized by strong mass
outflows around the equatorial plane where slow and dense disks will
be formed (Waters et al. 1988). The velocity profile of the
equatorial component can be described as
\begin{equation}
v_{\rm w,disk} (R) = v_{0,{\rm disk}} (\frac{R}{R_*})^m,
\end{equation}
where $R$ is the distance from the stellar surface, $v_{0,{\rm
disk}}$ is the wind velocity at the stellar surface, $R_*$ is the Be
star radius, $m$ is the outflow exponent. For a typical equatorial
outflow, $v_{0,{\rm disk}} \sim 10^6$ ${\rm cm}$ ${\rm s}^{-1}$, $m$
is in the range $0 < m < 2$. On the other hand, there is also a
polar wind component in the Be star outflow. Compared with the
equatorial component, the polar wind has much higher velocity but
lower density. The velocity distribution of the polar wind can be
approximated as (Waters et al. 1988)
\begin{equation}
v_{\rm w,polar} (R) = v_{0,{\rm polsar}}+(v_\infty-v_{0,{\rm
polar}})(1-R_*/R)^\beta,
\end{equation}
where $v_\infty$ is the terminal velocity of the wind at infinity
and $v_\infty \sim 100 v_{0,{\rm polsar}}$, the index $\beta$ is a
free parameter. For a typical Be star, $v_\infty \sim 10^8$ ${\rm
cm}$ ${\rm s}^{-1}$, and the value of $\beta$ is taken to be 1.5.

The dynamical pressure of the stellar wind can be described as
\begin{equation}
P_{\rm w} (R) = \rho_{\rm w} (R) {v_{\rm w} (R)}^2,
\end{equation}
where $\rho_{\rm w} (R) = \dot{M}/4 \pi f_{\rm w} R^2 v_w (R)$ is
the wind density, $\dot{M}$ is the mass-loss rate of the Be star,
$f_{\rm w}$ is the outflow fraction in units of $4\pi$ sr. For a
typical polar wind, $\dot{M}_{\rm polar}/f_{{\rm w},{\rm polar}}
\sim 10^{-8}$ $M_{\sun}$ ${\rm yr}^{-1}$; and for a typical
equatorial wind, $\dot{M}_{\rm disk}/f_{{\rm w},{\rm disk}} \sim
10^{-7}$ $M_{\sun}$ ${\rm yr}^{-1}$. We assumed in our model that
the density and velocity of the equatorial wind evolve with the
height above the equatorial plane in Gaussian forms to match the
polar wind. The dynamical pressure of the pulsar wind can be
described as
\begin{equation}
P_{\rm pul} (r) = \frac{L_{\rm sp}}{4 \pi f_{\rm p} c r^2},
\end{equation}
where $r$ is the distance from the pulsar, $L_{\rm sp}$ is the
spin-down luminosity of the pulsar, $f_{\rm p}$ is the pulsar wind
fraction in units of $4\pi$ sr and $c$ is the speed of light. We
assume that the pulsar wind is isotropic and $f_{\rm p} = 1$ in our
calculations. The location of the shock is determined by the
dynamical pressure balance between the pulsar wind and the stellar
wind, i.e.,
\begin{equation}
\frac{L_{\rm sp}}{4 \pi c r^2_{\rm s}} = \rho_{\rm w} (R_{\rm s})
{\vert\vec{v}_{\rm w} (R_{\rm s}) - \vec{v}_{\rm orb}\vert}^2,
\end{equation}
where $r_{\rm s}$ is the distance between the termination shock and
the pulsar, $R_{\rm s} = d - r_{\rm s}$ is the distance of the shock
from the stellar surface, $d$ is the separation between the pulsar
and its companion, $\vec{v}_{\rm orb}$ is the pulsar orbital
velocity (Tavani \& Arons 1997). The value and the direction of
$\vec{v}_{\rm orb}$ are changed along with the orbital phase.

\subsection{Magnetic field}

In the preshocked pulsar wind at $r_{\rm s}$, the magnetic field
$B_1$ and the proper electron number density $n_1$ can be described
as (Kennel \& Coroniti 1984a, 1984b)
\begin{equation}
B^2_1 = \frac{L_{\rm sp} \sigma}{r^2_{\rm s} c (1 + \sigma)},
\end{equation}
\begin{equation}
n_1 = \frac{L_{\rm sp}}{4 \pi u_1 \Gamma_1 r^2_{\rm s} m_{\rm e} c^3
(1 + \sigma)},
\end{equation}
where $u_1$ and $\Gamma_1 = \sqrt{1+u_1^2}$ are the dimensionless
radial four velocity and the bulk Lorentz factor of the unshocked
pulsar wind, $m_{\rm e}$ is the rest mass of the electron, $\sigma =
B^2_1/(4 \pi \Gamma_1 u_1 n_1 m_{\rm e} c^2)$ is the magnetization
parameter and is defined by the ratio of magnetic energy density and
particle kinetic energy density in the pulsar wind.

In the situation of the Crab Nebula, Kennel \& Coroniti (1984a,
1984b) took the magnetization parameter $\sigma \sim 0.003$ at a
distance of $r_{\rm s} \sim 3 \times 10^{17}$ cm from the pulsar. In
Tavani \& Arons (1997), the magnetization parameter was adopted as
$\sigma = 0.02$ for the PSR B1259-63/SS 2883 system. But at the
light cylinder, the typical value of $\sigma_{\rm L}$ should be as
large as
\begin{displaymath}
\sigma_{\rm L} = \frac{B^2_{\rm L}/8\pi}{2\dot{N}_{e^\pm} m_{\rm e} c/r_{\rm L}^2} \\
\end{displaymath}
\begin{equation}
\quad \sim 1.38 \times 10^7 (\frac{B_{\rm L}}{10^6 {\rm G}})^2
(\frac{r_{\rm L}}{10^8 {\rm cm}})^2 (\frac{N_{\rm m}}{10^4})^{-1}
(\frac{\dot{N}_{\rm GJ}}{5.26 \times 10^{31} {\rm s}^{-1}})^{-1},
\end{equation}
where $B_{\rm L}$ is the magnetic field at the light cylinder,
$r_{\rm L}$ is the radius of the light cylinder, $\dot{N}_{e^\pm} =
N_{\rm m} \dot{N}_{\rm GJ}$, $N_{\rm m}$ is the $e^\pm$ multiplicity
and $\dot{N}_{\rm GJ} \sim 5.26 \times 10^{31} (B/3 \times 10^{11}
{\rm G}) (P/47.762 {\rm ms})^{-2} {\rm s}^{-1}$ is the
Goldreich-Julian particle flow at the light cylinder. In outer gap
models (e.g. Cheng, Ho \& Ruderman 1986a, 1986b; Zhang \& Cheng
1997; Takata, Wang \& Cheng 2010), the multiplicity due to various
pair-creation processes could reach $10^4 - 10^5$. We can imagine
that between the light cylinder and the termination shock, the
magnetic energy will be gradually converted into the particle
kinetic energy. So in our work, we assume that $\sigma$ evolves with
$r_{\rm s}$. The profile of the variation used in our model can be
described as a power-law form,
\begin{equation}
\sigma = \sigma_{\rm L}(\frac{r}{r_{\rm L}})^{-\alpha_\sigma}.
\end{equation}
If we use the parameters of the Crab pulsar, i.e. $B_{\rm L} \sim
10^6$ G and $\dot{N}_{\rm GJ} \sim 10^{34}$ ${\rm s}^{-1}$, which
give $\sigma_{\rm L} \sim 10^5$ at $r_{\rm s} = 10^8$ cm, and
$\sigma = 0.003$ at $r_{\rm s} = 3 \times 10^{17}$ cm, then the
typical value of $\alpha_\sigma$ is of order of unity. Note that
here we obtain $\alpha_\sigma$ by comparing with the Crab pulsar and
assuming the conversion efficiency from the magnetic energy to the
particle kinetic energy is a simple power-law. But the actual
condition in the PSR B1259-63/SS 2883 system may not be the same and
larger or smaller values of $\alpha_\sigma$ may also be possible. So
the exact value of $\alpha_\sigma$ is defined as a parameter in our
calculations. According to the energy conservation, $(1+\sigma)
\Gamma_1 \sim constant$, we define $\Gamma_0$ as the value of
$\Gamma_1$ when $\sigma = 0$ and take it as a free parameter. We can
see that $\Gamma_1$ also evolves with $r_{\rm s}$ due to the energy
conservation.

The downstream magnetic field is described as (Kennel \& Coroniti
1984a, 1984b)
\begin{equation}
B_2=B_1\sqrt{1+\frac{1}{u^2_2}},
\end{equation}
\begin{equation}
u^2_2 =
\frac{8\sigma^2+10\sigma+1}{16(\sigma+1)}+\frac{[64\sigma^2(\sigma+1)^2+20\sigma(\sigma+1)+1]^\frac{1}{2}}{16(\sigma+1)}.
\end{equation}

\subsection{Electron distribution}

It is usually assumed that the unshocked cold electron pairs can be
accelerated to a power-law distribution in the termination shock
front, and be injected into the downstream post-shock flow. So we
adopt a power-law electron injection spectrum $\dot{Q}(\gamma_{\rm
e}) \sim (\gamma_{\rm e}-1)^{-p}$ ($\gamma_{\rm e,min} < \gamma_{\rm
e} < \gamma_{\rm e,max}$) in our work, where $p$ is the electron
distribution index and usually varies between 2 and 3. Since the
shock acceleration is a highly nonlinear process, it is very
difficult to know how $p$ evolves in different orbital phase. For
simplicity, we assume that $p$ is a constant. The minimum Lorentz
factor can be determined from the conservations of the total
electron number $L_{\rm sp}/\Gamma m_{\rm e} c^2=\int
\dot{Q}(\gamma_{\rm e}) {\rm d} \gamma_{\rm e}$ and the total
electrons energy $L_{\rm sp}=\int \dot{Q}(\gamma_{\rm e})
\gamma_{\rm e} m_{\rm e} c^2 {\rm d} \gamma_{\rm e}$ (Kirk, Ball \&
Skj{\ae}raasen 1999), and we can acquire
\begin{equation}
\gamma_{\rm e,min}=\Gamma_1\frac{p-2}{p-1}.
\end{equation}
The maximum Lorentz factor of electrons can be determined by
equating the cooling timescale of electrons $t_{\rm cool}$ with the
particle acceleration timescale $t_{\rm ac}$ as the following form,
\begin{equation}
\gamma_{\rm e,max}=\sqrt{\frac{6 \pi e}{\sigma_{\rm T} B_2 (1+Y)}},
\end{equation}
where $e$ is the electron charge, $\sigma_{\rm T}$ is the Thompson
scattering cross section, $Y$ is the Compton parameter and is
defined below. The cooling timescale $t_{\rm cool} = \gamma_{\rm e}
m_{\rm e} c^2 / P_{\rm rad}$, where $P_{\rm rad} = (4/3) \sigma_{\rm
T} c \gamma_{\rm e}^2 (B^2_2/8\pi) (1+Y)$ is the radiation power of
electrons. The particle acceleration timescale $t_{\rm ac}=
\gamma_{\rm e} m_{\rm e} c/e B_2$.

The cooling of electrons due to synchrotron and IC radiation will
steepen the distribution of electrons above a critical Lorentz
factor $\gamma_{\rm e,c}$, which can be derived from equating the
cooling timescale $t_{\rm cool}$ with the dynamic flow time
$\tau_{\rm dyn}$ of electrons. So $\gamma_{\rm e,c}$ can be
expressed as (Sari, Piran \& Narayan 1998)
\begin{equation}
\gamma_{\rm e,c}=\frac{6 \pi m_{\rm e} c}{\sigma_{\rm T} B^2_2
\tau_{\rm dyn}(1+Y)}.
\end{equation}
In our model, we assume the dynamic flow time of electrons in the
radiation cavity is $\tau_{\rm dyn} = \xi r_{\rm s} / v_{\rm f}$,
where $\xi$ is a dimensionless parameter and $v_{\rm f}$ is the
postshock flow velocity. In Tavani \& Arons (1997), they used $\xi =
3$ by taking into account the nonspherical shape of the shocked
region and $v_{\rm f} = c/3$. The electrons with Lorentz factor
$\gamma_{\rm e}
> \gamma_{\rm e,c}$ will be cooled by the radiation rapidly.

The Compton parameter $Y$ used above describes the effect of IC
(Sari \& Esin 2001), which is defined as the ratio of the IC
luminosity (including SSC and EIC) to the synchrotron luminosity,
\begin{displaymath}
Y(\gamma_{\rm e}) \equiv \frac{L_{\rm SSC}+L_{\rm EIC}}{L_{\rm SYN}}
\end{displaymath}
\begin{equation}
\qquad \quad =\frac{u_{\rm syn}[\nu \leq \nu_{\rm KN}(\gamma_{\rm
e})]+u_{\rm star}[\nu \leq \nu_{\rm KN}(\gamma_{\rm e})]}{u_B},
\end{equation}
\begin{equation}
u_{\rm syn}[\nu \leq \nu_{\rm KN}(\gamma_{\rm
e})]=\frac{\eta(\gamma_{\rm e})}{1+Y(\gamma_{\rm e})} \frac{L_{\rm
sp}}{4 \pi c r^2_{\rm s}},
\end{equation}
\begin{equation}
u_{\rm star}[\nu \leq \nu_{\rm KN}(\gamma_{\rm e})]=\frac{L_{\rm
star}[\nu \leq \nu_{\rm KN}(\gamma_{\rm e})]}{4 \pi c (d-r_{\rm
s})^2},
\end{equation}
\begin{equation}
u_B=\frac{B^2_2}{8 \pi},
\end{equation}
where $\nu_{\rm KN}(\gamma_{\rm e}) = m_{\rm e} c^2/(h \gamma_{\rm
e})$ is the critical frequency of scattering photons above which the
scattering with electrons of Lorentz factor $\gamma_{\rm e}$ enter
the Klein-Nishina (KN) regime, $\eta(\gamma_{\rm e})=\eta_{\rm rad}
\eta_{\rm KN}(\gamma_{\rm e})$ is the radiation efficiency,
$\eta_{\rm rad}$ is the fraction of the electron's energy which is
radiated away, and $\eta_{\rm KN}(\gamma_{\rm e})$ is the fraction
of synchrotron photons below the KN limit frequency $\nu_{\rm KN}
(\gamma_{\rm e})$. In the slow cooling case, $\gamma_{\rm e,min} <
\gamma_{\rm e,c}$, $\eta_{\rm rad} = (\gamma_{\rm e,c}/\gamma_{\rm
e,min})^{2-p}$ (Wang et al. 2010),
\begin{equation}
\eta_{\rm KN}(\gamma_{\rm e}) = \left \{
\begin{array}{ll}
 (\frac{\nu_{\rm m}}{\nu_{\rm c}})^\frac{3-p}{2} (\frac{\nu_{\rm KN}(\gamma_{\rm e})}{\nu_{\rm m}})^\frac{4}{3}, & \nu_{\rm KN}(\gamma_{\rm e}) \leq \nu_{\rm m}, \\
 (\frac{\nu_{\rm KN}(\gamma_{\rm e})}{\nu_{\rm c}})^\frac{3-p}{2}, & \nu_{\rm m} < \nu_{\rm KN}(\gamma_{\rm e}) < \nu_{\rm c}, \\
 1, & \nu_{\rm c} \leq \nu_{\rm KN}(\gamma_{\rm e}); \\
 \end{array} \right.
\end{equation}
In the fast cooling case, $\gamma_{\rm e,c} < \gamma_{\rm e,min}$
,$\eta_{\rm rad} = 1$ (Wang et al. 2010),
\begin{equation}
\eta_{\rm KN}(\gamma_{\rm e}) = \left \{
\begin{array}{ll}
 (\frac{\nu_{\rm c}}{\nu_{\rm m}})^\frac{1}{2} (\frac{\nu_{\rm KN}(\gamma_{\rm e})}{\nu_{\rm c}})^\frac{4}{3}, & \nu_{\rm KN}(\gamma_{\rm e}) \leq \nu_{\rm c}, \\
 (\frac{\nu_{\rm KN(\gamma_{\rm e})}}{\nu_{\rm m}})^\frac{1}{2}, & \nu_{\rm c} < \nu_{\rm KN}(\gamma_{\rm e}) < \nu_{\rm m}, \\
 1, & \nu_{\rm m} \leq \nu_{\rm KN}(\gamma_{\rm e}), \\
 \end{array} \right.
\end{equation}
where $\nu_{\rm m}$ and $\nu_{\rm c}$ are the typical synchrotron
radiation frequencies of electrons whose Lorentz factors are
$\gamma_{\rm e,min}$ and $\gamma_{\rm e,c}$ respectively, and are
defined as $\nu_{\rm m} = 3 \gamma_{\rm e,m}^2 e B/(4 \pi m_{\rm e}
c)$ and $\nu_{\rm c} = 3 \gamma_{\rm e,c}^2 e B/(4 \pi m_{\rm e} c)$
respectively.

The radiation revised electron spectrum $N(\gamma_{\rm e})$ can be
obtained from the continuity equation of the electron distribution
(Ginzburg \& Syrovatshii 1964),
\begin{equation}
\frac{\partial n(\gamma_{\rm e},t)}{\partial t}+\frac{\partial
\dot{\gamma}_{\rm e} n(\gamma_{\rm e},t)}{\partial \gamma_{\rm
e}}=\dot{Q}(\gamma_{\rm e}).
\end{equation}
Similar to the LS I+61\textordmasculine 303 system, because the
cooling and dynamic flow timescales are much smaller than the
orbital period in the PSR B1259-63/SS 2883 system, we use $\partial
n(\gamma_{\rm e},t)/\partial t = 0$ to calculate the electron
distribution at a steady state (Zabalza, Parades \& Bosch-Ramon
2011) and obtain,\\
(1) $\gamma_{\rm e,c} \le \gamma_{\rm e,min} < \gamma_{\rm e,max}$,
\begin{equation}
N(\gamma_{\rm e}) \propto \left\{
\begin{array}{ll}
 \left( \gamma _{\rm e} - 1 \right)^{- 2}, & \gamma _{\rm e,c}  \le \gamma _{\rm e}  < \gamma _{\rm e,min }, \\
 \left( \gamma _{\rm e} - 1 \right)^{ -(p + 1)}, & \gamma _{\rm e,min }  \le \gamma _{\rm e}  \le \gamma _{\rm e,max }; \\
 \end{array} \right.
\end{equation}
(2) $\gamma_{\rm e,min} < \gamma _{\rm e,c} \le \gamma _{\rm
e,max}$,
\begin{equation}
N(\gamma_{\rm e}) \propto \left\{
\begin{array}{ll}
 \left( \gamma _{\rm e} - 1 \right)^{ - p}, & \gamma _{\rm e,min }  \le \gamma _{\rm e}  \le \gamma _{\rm e,c}, \\
 \left( \gamma _{\rm e} - 1 \right)^{ -(p + 1)}, & \gamma _{\rm e,c}  < \gamma _{\rm e}  \le \gamma _{\rm e,max }; \\
 \end{array} \right.
\end{equation}
(3) $\gamma _{\rm e,c}  > \gamma _{\rm e,max}$,
\begin{equation}
N(\gamma_{\rm e}) \propto ( \gamma _{\rm e} - 1)^{-p},  \gamma _{\rm
e,\min } \le \gamma _{\rm e}  \le \gamma _{\rm e,\max}.
\end{equation}
The coefficient can be calculated from $N_{\rm tot} = \int
N(\gamma_{\rm e}) {\rm d} \gamma_{\rm e}$, where $N_{\rm tot} =
(\Delta\Omega/4\pi) L_{\rm sp} \tau_{\rm dyn}/[\Gamma_1 m_{\rm e}
c^2 (1+\sigma)]$, and $\Delta\Omega/4\pi$ is the fraction of pulsar
wind electrons which are accelerated to relativistic velocities in
the shock front. In our calculations, we assume that the typical
scale of the radiation cavity is $r_{\rm s}$ for simplicity, and
therefore $\Delta\Omega \sim \pi/4$.

\subsection{Radiation process}

In our model, the multi-band photons of PSR B1259-63/SS 2883 are
from the synchrotron radiation and IC radiation (including SSC
scattering of the synchrotron photons and EIC scattering of the
thermal photons from the Be star) of the shock-accelerated
electrons.

The synchrotron radiation power at frequency $\nu$ from a single
electron with Lorentz factor $\gamma_{\rm e}$ is given by (Rybicki
\& Lightman 1979)
\begin{equation}
P^{\rm SYN}_\nu(\gamma_{\rm e}) = \frac{\sqrt{3} e^3 B}{m_{\rm e}
c^2} F (\frac{\nu}{\nu_{\rm c}}),
\end{equation}
where $\nu_{\rm c} = 3 \gamma_{\rm e}^2 e B/(4 \pi m_{\rm e} c)$.
The function $F(x)$ is defined as
\begin{equation}
F(x) = x \int_{x}^{+ \infty} K_{5/3}(k) {\rm d}k.
\end{equation}
The SSC radiation power at frequency $\nu$ from a single electron
with Lorentz factor $\gamma_{\rm e}$ is given by (Blumenthal \&
Gould 1970)
\begin{equation}
P^{\rm SSC}_\nu(\gamma_e) = 3 \sigma_{\rm T} \int^\infty_{\nu_{\rm
s,min}} {\rm d} \nu_{\rm s} \frac{\nu f_{\nu_{\rm s}}^{\rm SYN}}{4
\gamma^2_e \nu^2_{\rm s}} g(x,y),
\end{equation}
\begin{equation}
g(x,y) = 2y{\rm ln}y+(1+2y)(1-y)+\frac{x^2y^2}{2(1+xy)}(1-y),
\end{equation}
where $f_{\nu_{\rm s}}^{\rm SYN} = \int P^{\rm SYN}_{\nu_{\rm
s}}(\gamma_{\rm e}) N(\gamma_{\rm e}) {\rm d} \gamma_{\rm e}/4 \pi
r^2_{\rm s}$ is the flux density of the synchrotron radiation,
$x=4\gamma_{\rm e}h\nu_{\rm s}/(m_{\rm e}c^2)$,
$y=h\nu/[x(\gamma_{\rm e}m_{\rm e}c^2-h\nu)]$, $\nu_{\rm s,min}=\nu
m_{\rm e}c^2/[4\gamma_{\rm e}(\gamma_{\rm e}m_{\rm e}c^2-h\nu)]$.
Correspondingly, the EIC radiation power at frequency $\nu$ from a
single electron with Lorentz factor $\gamma_{\rm e}$ is given by
(Aharonian \& Atoyan 1981; He et al. 2009)
\begin{equation}
P^{\rm EIC}_\nu(\gamma_e,{\rm cos}\theta_{\rm SC}) = 3 \sigma_{\rm
T} \int^\infty_{\nu_{\rm s, min}} {\rm d} \nu_{\rm s} \frac{\nu
f_{\nu_{\rm s}}^{\rm STAR}}{4 \gamma^2_e \nu^2_{\rm s}}
h(\xi,b_\theta),
\end{equation}
\begin{equation}
h(\xi,b_\theta) =
1+\frac{\xi^2}{2(1-\xi)}-\frac{2\xi}{b_\theta(1-\xi)}+\frac{2\xi^2}{b^2_\theta(1-\xi)^2},
\end{equation}
where $f_{\nu_{\rm s}}^{\rm STAR} = \pi B_\nu(T_{\rm
eff})(R_*/R_{\rm s})^2$ is the flux density of the Be star photons,
$B_\nu(T_{\rm eff}) = 2h\nu^3/c^2[{\rm exp}(h\nu/kT_{\rm eff})-1]$
is the brightness on the Be star surface, $T_{\rm eff}$ is the
effective temperature of the star, $h$ is the Planck constant, $k$
is the Boltzmann constant, $\sigma_{\rm T}$ is the Thompson
cross-section, $\xi=h\nu/(\gamma_{\rm e}m_{\rm e}c^2)$,
$b_\theta=2(1-{\rm cos}\theta_{\rm SC})\gamma_{\rm e}h\nu_{\rm
s}/(m_{\rm e}c^2)$, $h\nu_{\rm s}\ll h\nu \leq \gamma_{\rm e}m_{\rm
e}c^2b_\theta/(1+b_\theta)$, $\theta_{\rm SC}$ is the angle between
the injecting photons and the scattered photons, and is varied along
with the orbital phase. Note that here the Be star is assumed
as a point-like and black body emitter for simplicity.

The observed total flux densities from all the shock accelerated
electrons at frequency $\nu$ is
\begin{displaymath}
F_{\nu} ({\rm cos}\theta_{\rm SC}) = \frac{1}{4 \pi D_{\rm L}^2}
\int^{\gamma_{e, {\rm max}}}_{{\rm min}(\gamma_{e, {\rm
c}},\gamma_{e, {\rm min}})} {\rm d} \gamma_{\rm e} N(\gamma_{\rm e})
[P^{\rm SYN}_\nu (\gamma_e)
\end{displaymath}
\begin{equation}
\qquad \qquad \qquad +P^{\rm SSC}_\nu (\gamma_e)+P^{\rm EIC}_\nu
(\gamma_e,{\rm cos}\theta_{\rm SC})].
\end{equation}
As the pulsar orbiting around the Be star, the spatial condition and
the parameter ${\rm cos}\theta_{\rm SC}$ will be changed. Some
physical conditions, such as the the electron distribution
$N(\gamma_{\rm e})$, the downstream magnetic field and the photon
field, will vary accordingly. So a variable flux along with the
orbital phase will appear.

The gamma-ray photons produced by the IC process could be
absorbed in the dense stellar photon field through pair production
(Gould \& Schr\'eder 1967). The opacity could be roughly estimated
by $\tau_{\gamma \gamma} \approx \sigma_{\gamma \gamma} n_* R_{\rm
s} \approx 0.67$ at periastron, where $\sigma_{\gamma \gamma}
\approx \sigma_{\rm T}/5$, $n_*$ is the stellar photon density.
Because the opacity at the periastron is already the largest value 
in all orbital phases but
it is still less than unity, we will ignore the effect of
pair production in our calculations for simplicity.

\section{Results}

In this section, we will present some calculated results using our
model, and compare them with the observations. Some parameters used
in our calculation for the PSR B1259-63/SS 2883 system are as
follows (Tavani \& Arons 1997): For the orbital parameters, we take
the eccentricity $e = 0.87$, the semimajor axis $a = 4.8$ AU; For
the compact object PSR B1259-63, we take the spin-down luminosity
$L_{\rm sp} = 8 \times 10^{35}$ $\rm erg$ $\rm s^{-1}$; For the Be
star SS 2883, we take the stellar luminosity $L_{\rm star} = 5.8
\times 10^4 L_\odot$, the stellar mass $M_{\rm star} = 10 M_\odot$,
the stellar radius $R_* = 10 R_\odot$, the effective temperature of
the star $T_{\rm eff} = 27000$ $\rm K$. The distance between the
system and Earth is taken to be 1.5 kpc (Johnston et al. 1994). We
use $\xi = 3$ and $v_{\rm f} = c/3$ in our calculations, which are
the same as those suggested by Tavani \& Arons (1997).

\subsection{Variation of some shock parameters}

\begin{figure}
\resizebox{\hsize}{!}{\includegraphics{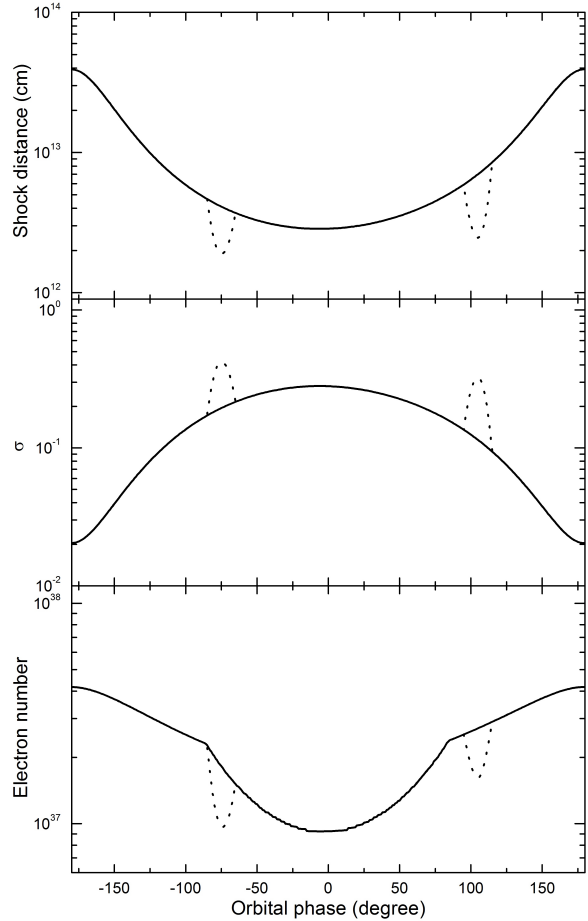}} \caption{Top panel:
The variation of shock distance $r_{\rm s}$ along with the orbital
phase; Middle panel: The variation of magnetization parameter
$\sigma$ along with the orbital phase; Bottom panel: The variation
of electron number $N_{\rm 1-10 keV}$ along with the orbital phase.
The solid and dotted lines correspond to the conditions without and
with the effect of disk respectively.}
\end{figure}

We will first give some results on the variations of the distance
between the shock and the pulsar $r_{\rm s}$, the magnetization
parameter $\sigma$ and the electron number $N_{\rm 1-10 keV}$ whose
typical synchrotron emission frequencies are in the range of 1-10
keV as a function of orbital phase, as shown in Figure 1. The
periastron is taken to be at orbital phase $\sim 0\textordmasculine
$ throughout the paper. In solid lines, we only consider the polar
wind and the input parameters we used are as follows: the terminal
velocity of the polar wind $v_\infty = 1.3 \times 10^8$ ${\rm cm}$
${\rm s}^{-1}$, the index of the polar wind $\beta = 1.5$,
$\dot{M}_{\rm polar}/f_{{\rm w},{\rm polar}} = 2 \times 10^{-8}$
$M_{\sun}$ ${\rm yr}^{-1}$, the electron distribution index $p=2.2$,
the magnetization parameter at light cylinder ($r_{\rm s} = 10^8
{\rm cm}$) $\sigma_{\rm L} = 8 \times 10^3$, the index
$\alpha_\sigma = 1.0$ and $\Gamma_0 = 2 \times 10^5$. In dotted
lines, we add an equatorial component with the initial velocity
$v_{\rm 0,disk} = 10^6$ ${\rm cm}$ ${\rm s}^{-1}$, the index $m =
1.5$ and $\dot{M}_{\rm disk}/f_{{\rm w},{\rm disk}} = 2 \times
10^{-7}$ $M_{\sun}$ ${\rm yr}^{-1}$. In our calculations, the
half-opening angle of the disk (projected on the pulsar orbital
plane) is $\Delta \theta_{\rm disk} = 10\textordmasculine$, and the
intersection between the stellar equatorial plane and the orbital
plane is inclined at $\theta_{\rm disk} = 75\textordmasculine$ to
the major axis of the pulsar orbit, which are in the range suggested
by Chernyakova et al. (2006) with $\Delta \theta_{\rm disk} \simeq
18\textordmasculine.5$ and $\theta_{\rm disk} \simeq
70\textordmasculine$. These input parameters we used here are chosen
by modeling the observations (See below). It can be seen that
because the shock distance $r_{\rm s}$ varies along with the orbital
phase, the magnetization parameter $\sigma$ and the electron number
$N_{\rm 1-10 keV}$ evolve accordingly. We can also see from the
bottom panel of Figure 1 that because the total electron number
evolves as $N_{\rm tot} \propto r_{\rm s}$, $N_{\rm 1-10 keV}$
varies as $N_{\rm 1-10 keV} \propto r_{\rm s}$ when $r_{\rm s}$ is
large. As the pulsar approaching the periastron, the IC process
is more and more important and the value of $Y(\gamma_{e,c})$ will
increase. At a certain position, some electrons will enter the fast
cooling case ($\gamma_{\rm e,c} < \gamma_{\rm e,min}$) with the
distribution index being $-(p+1)$ instead of $-p$ in the slow
cooling case. At this moment, the electrons will be distributed in a
more broad range, and the value of $N_{\rm 1-10 keV}$ will decrease
more rapidly around the periastron because more and more electrons
are cooled to the lower energies. In the dense disk, $r_{\rm s}$
and $\sigma$ reach a minimum and a maximum respectively, and $N_{\rm
1-10 keV}$ reaches a minimum because of the smallest electron number
and the fastest cooling effect.

\subsection{Spectrum}

The calculated spectra from $10^{-2}$ eV to 100 TeV are illustrated
in Figure 2. The input parameters used here are the same as those in
Figure 1. The different types of lines represent the spectra in
different orbital phases: the solid line and the dashed line
correspond to the conditions at periastron and at the orbital phase
of $100\textordmasculine$ respectively, where the effect of the disk
is not considered; the dotted line corresponds to the result in the
disk (the orbital phase = $-75\textordmasculine$).

\begin{figure}
\resizebox{\hsize}{!}{\includegraphics{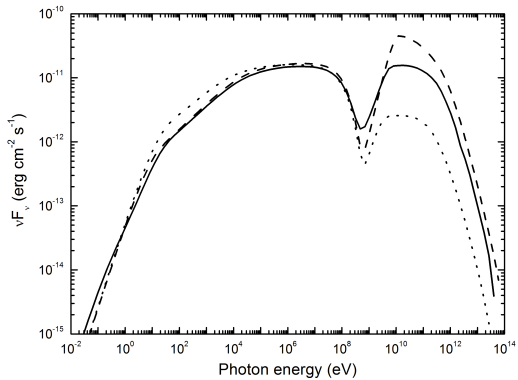}} \caption{The
calculated spectra. The solid line and the dashed line correspond to
the conditions at periastron and at the orbital phase of
$100\textordmasculine$ respectively, where the effect of disk is not
considered; the dotted line corresponds to the result in the disk.}
\end{figure}

The emission below 100 MeV is mainly from the synchrotron radiation.
The shape of synchrotron spectrum is mainly determined by the
magnetic field and the electron distribution. It can be seen that
between 1-10 keV the flux in the dotted line is the highest in all
the three lines. Although the electron number $N_{\rm 1-10 keV}$
reaches a minimum in the disk, the radiation flux can be very large
because the magnetic field is highest at this moment, as shown in
Figure 1. The large flux in the disk will produce an ``emission''fa
profile in the X-ray light curve. We can also see that in solid line
the spectrum below 10 eV is softer and the flux below 1 eV is higher
than those in other two lines, which is because more electrons are
cooled to this energy range at periastron.

The emission above 1 GeV is mainly produced by the EIC process. The
calculated SSC flux is lower than the EIC flux by more than three
orders of magnitude for the reason that the SSC emission is
strongly suppressed by the KN effect, so we neglect the effect of
SSC in our calculations throughout the paper. There are two breaks
in the EIC spectra corresponding to the two critical energies
$\gamma_{\rm e,min} m_{\rm e} c^2$ and $(m_{\rm e} c^2)^2/k T_{\rm
eff}$ (Yu 2009). In our calculations, $\gamma_{\rm e,min} m_{\rm e}
c^2$ corresponds to the first break and $(m_{\rm e} c^2)^2/k T_{\rm
eff} \sim$ 0.1 TeV corresponds to the second break in the EIC
spectra. We can see that the flux in the dotted line is the lowest
in all the three lines. The reasons are as follows: (1) In the disk,
the magnetization parameter $\sigma$ gets a maximum and $\gamma_{\rm
e,min}$ reaches a minimum accordingly. The small $\gamma_{\rm
e,min}$ makes the EIC spectrum begin to break at a low energy; (2)
the electron number whose EIC emission frequencies are above 1 GeV
decreases duo to the rapid cooling in the disk; (3) The strong wind
in the disk pushes the shock to the pulsar, so the photon density
from Be star decrease. The low flux in the disk will produce an
``absorption'' profile in the TeV light curve.

The $Fermi$ Gamma-ray Space Telescope observed the PSR B1259-63/SS
2883 system recently (Tam et al. 2010; Abdo et al. 2010a, 2010b,
2011b; Kong et al. 2011). The Large Area Telescope (LAT) onboard
$Fermi$ covers the energy band from 20 MeV to greater than 300 GeV.
We can see that in our model, the emission in this energy range is
produced by the synchrotron and the EIC processes together (See
Section 3.6 for detailed discussion).

\subsection{Comparison with observations in the 1-10 keV band}

The X-rays from the PSR B1259-63/SS 2883 system was first detected
by ROSAT (Cominsky, Roberts \& Johnston 1994). After that, the ASCA
satellite also observed this system in X-rays (Kaspi et al. 1995;
Hirayama et al. 1999). The results from the ASCA satellite show that
the X-ray flux from the source is highly variable in different
orbital phase, and the light curve has two peaks before and after
the periastron respectively. More recently, new X-ray observation
data of this source were reported by Chernyakova et al. (2006,
2009). The results include the $XMM-Newton$ observations at the
beginning of 2004, some earlier unpublished X-ray data from
$BeppoSAX$ and $XMM-Newton$ (Chernyakova et al. 2006), and the
unprecedented detailed observations by $Suzaku$, $Swift$,
$XMM-Newton$ and $Chandra$ missions during the 2007 periastron
passage (Chernyakova et al. 2009). These new data are well in
agreement with previous observations, and show that the orbital
light curve in X-rays does not exhibit strong orbit-to-orbit
variations (Chernyakova et al. 2009).

\begin{figure}
\resizebox{\hsize}{!}{\includegraphics{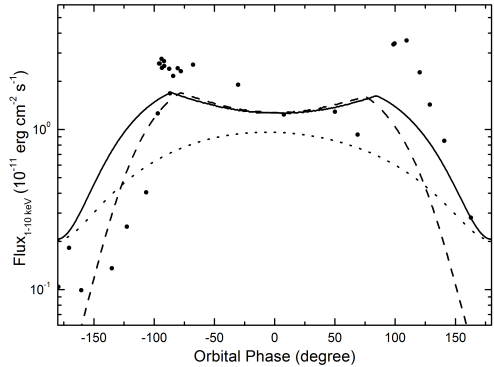}} \caption{The
calculated X-ray light curves and their comparison with
observations. The solid line and the dashed line correspond to
$\alpha_\sigma = 1.0$ and $\alpha_\sigma = 2.0$ respectively. The
dotted line corresponds to a constant magnetization parameter
$\sigma = 0.02$.}
\end{figure}

Here, we use our model described in Section 2 to reproduce the X-ray
light curve of the PSR B1259-63/SS 2883 system. We assume that the
X-ray emission is mainly from the synchrotron radiation of the
relativistic electrons. The X-ray data are taken from Chernyakova et
al. (2006, 2009). In this subsection, we will only use the polar
component as the stellar wind. The role of the disk will be
discussed later. The calculated light curves and the observed data
in 1-10 keV band are presented in Figure 3. Except for special
declaration, the parameters we used here are the same as those in
the solid lines of Figure 1. In the dotted line, we use a constant
value of $\sigma = 0.02$ for reference purpose. Differently, in the
solid and dashed lines, we make the magnetization parameters vary as
$\sigma = \sigma_{\rm L}(r/r_{\rm L})^{-\alpha_\sigma}$. We can see
that when we use a constant magnetization factor $\sigma = 0.02$, as
suggested by Tavani \& Arons (1997), the resulting light curve
cannot explain the two major characteristics in observations: (1)
The observed 1-10 keV flux has a sharp rise of about 20 times from
apastron to pre-periastron, but the ratio of flux between the
maximum and the minimum is only about 5 in the dotted line; (2) In
the dotted line, the flux get the maximum at periastron, in
opposition to the observations where the flux has a minimum around
the periastron. When the variation of $\sigma$ is considered, the
calculated results fit the above two characteristics much better.
The appearance of this type of bimodal structure is due to the joint
action of the variable magnetic field and electron number. As shown
in Figure 1, as the pulsar approaching the periastron from the
apastron, the magnetization parameter $\sigma$ becomes larger, so
the magnetic field $B$ and the synchrotron radiation flux in X-rays
($L_{\rm X}$) should increase more rapidly than those in the
situation where $\sigma$ is a constant. A much sharper rise of about
20 times in flux could be produced at this moment. On the other
hand, a larger magnetic field $B$ and a smaller shock distance
$r_{\rm s}$ around the periastron will make the cooling of particles
stronger. Much more electrons will be cooled to the lower energies,
and the electron number $N_{1-10 \rm keV}$ decreases more rapidly
accordingly, as shown in Figure 1. So the flux in the 1-10 keV range
has a minimum around the periastron.

The solid and dashed lines in Figure 3 illustrate the effect of the
index $\alpha_\sigma$ on the 1-10 keV light curves. The parameters
used in solid line are the same as those used in the solid line of
Figure 1. We choose the value of the magnetization parameter
$\sigma$ as a constant at periastron, and take $\alpha_\sigma = 2.0$
in the dashed line. It is clearly seen that when $\alpha_\sigma$
becomes larger, the rise and the drop of flux around the apastron
become sharper. This is due to the more rapid change in the value of
$\sigma$.

\subsection{Comparison with VHE observations above 1 TeV band}

As early as 1999, Kirk, Ball \& Skj{\ae}raasen (1999) predicted that
the PSR B1259-63/SS 2883 system could produce VHE radiation, and the
light curve should reach the maximum around the periastron. About 5
years later, the TeV $\gamma$-rays from this source were first
detected by the High Energy Stereoscopic System (HESS) around its
periastron passage in 2004 (Aharonian et al. 2005). This makes the
PSR B1259-63/SS 2883 system be the first known binary system to emit
photons above 1 TeV (Aharonian et al. 2005). The observed TeV light
curve is similar to that in 1-10 keV for having two peaks in pre-
and post-periastron respectively with a minimum in flux around the
periastron (Aharonian et al. 2005), and is contrary to the
prediction by Kirk, Ball \& Skj{\ae}raasen (1999). The TeV light
curve was also found to have significant variations on timescales of
days, which makes the system be the first variable VHE source in our
Galaxy (Aharonian et al. 2005). The HESS team also observed the
source in 2005, 2006 and around the 2007 periastron passage
(Aharonian et al. 2009). They found that there is no significant VHE
$\gamma$-ray flux near apastron in the 2005 and 2006 observations,
and confirmed that this system is a variable TeV emitter.

Here, we use our model introduced in Section 2 to reproduce the TeV
light curve of the PSR B1259-63/SS 2883 system. We consider that the
photons above 1 TeV are mainly produced by the EIC process. Our
modelling light curve are presented in Figure 4 and the parameters
used here are the same as those used in Figure 3. The observational
TeV data are taken from Aharonian et al. (2005, 2009). We can see
that our model can explain the two-peak profile in TeV observations
qualitatively.

\begin{figure}
\resizebox{\hsize}{!}{\includegraphics{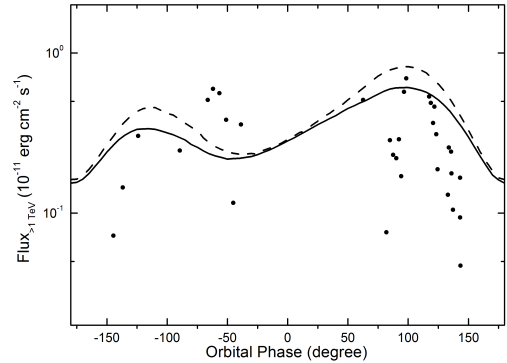}} \caption{The
calculated TeV light curves and their comparison with observations.
The solid line and the dashed line correspond to $\alpha_\sigma =
1.0$ and $\alpha_\sigma = 2.0$ respectively.}
\end{figure}

\subsection{Effect of the equatorial component in the stellar wind}

Be stars like SS 2883 are characterized by strong mass outflows
around the equatorial plane. They can produce disks which are slower
and denser than the polar wind (Waters et al. 1988). The existence
of a disk in the PSR B1259-63/SS 2883 system has been confirmed by
the radio observations (Johnston et al. 1996, 2005). The disk of Be
star SS 2883 in this binary system is believed to be tilted with
respect to the orbital plane. The line of intersection between the
disk plane and the orbital plane is oriented at about
90\textordmasculine with respect to the major axis of the binary
orbit, but the inclination of the disk is not constrained (Wex et
al. 1998; Wang, Johnston \& Manchester 2004). Chernyakova et al.
(2006) analyzed the X-ray light curve of the system detailedly and
suggested that the disk has a well-defined Gaussian-profile. The
half-opening angle of the disk (projected on the pulsar orbital
plane) is $\Delta \theta_{\rm disk} \simeq 18\textordmasculine.5$,
and the intersection between the stellar equatorial plane and the
orbital plane is inclined at $\theta_{\rm disk} \simeq
70\textordmasculine$ to the major axis of the pulsar orbit.

The presence of disk not only affects the radio radiation from the
system, but also plays an important role in producing the X-ray and
VHE $\gamma$-ray emission (Chernyakova et al. 2006). Ball et al.
(1999) suggested that the acceleration process will be most
efficient when the pulsar passes through the disk. Kawachi et al.
(2004) suggested that in addition to the relativistic electrons, VHE
protons could be produced by the interaction between the pulsar wind
and the dense equatorial wind. These hypothesis predict that there
will be an increase of flux in disk in both the X-ray and TeV bands
and can explain the two-peak profiles in the observed light curves.
Recently, Kerschhaggl (2010) found that there is a drop in the TeV
light curve when the pulsar passes through the dense disk. They
suggested that this is due to an increasing nonradiative cooling
effect in the disk.

Here we add the equatorial component of the stellar wind in our
calculations, and assume the mass-loss rate and velocity of the
equatorial wind have Gaussian profiles to match those of the polar
wind. Our results in the X-ray and TeV bands are presented in Figure
5. The solid and dotted lines correspond to the conditions with and
without equatorial components respectively. The parameters used here
are the same as those used in Figure 1. We can see in Figure 5 that
the X-ray flux increases in the passage of disk, but the flux in TeV
range decreases significantly at this moment, which is consistent
with the analysis by Kerschhaggl (2010). As discussed in Section
3.2, the appearance of an emission component in the X-ray light
curve is mainly due to the competition between the decrease of the
electron number and the increase of the magnetic field inside the
disk, whereas the appearance of an absorption component in the TeV
range is due to the low break energy, the decrease of electron
number resulting from rapid cooling and the decrease of photon
density in the disk.

\begin{figure}
\resizebox{\hsize}{!}{\includegraphics{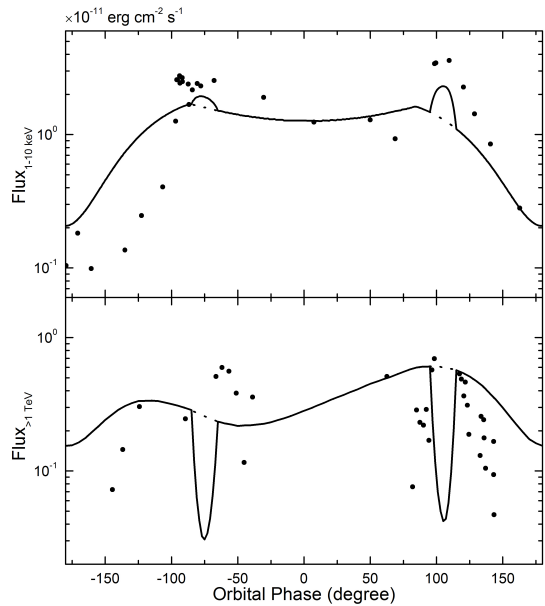}} \caption{The effect
of disk on the X-ray and TeV light curves. The solid and dotted
lines correspond to the conditions with and without the effect of
disk respectively.}
\end{figure}

\subsection{Observations by LAT onboard $Fermi$}

The PSR B1259-63/SS 2883 system passed through its periastron in
mid-December 2010 recently. The LAT onboard $Fermi$ observed this
source detailedly. An average flux (above 100 MeV) of $1.2 \pm 0.6
\times 10^{-7}$ $\rm ph$ $\rm cm^{-2}$ $\rm s^{-1}$ and a photon
index of $2.2 \pm 0.2$ in the period 2010-11-17 to 2010-12-19 UTC
was reported by the $Fermi$ team (Abdo et al. 2010b). They also
reported a flux (above 100 MeV) of $1.7 \pm 0.4 \times 10^{-6}$ $\rm
ph$ $\rm cm^{-2}$ $\rm s^{-1}$ with a photon index of $3.0 \pm 0.3$
in the period 2011-01-17 to 2011-01-19 UTC, and a flux (above 100
MeV) of $1.8 \pm 0.2 \times 10^{-6}$ $\rm ph$ $\rm cm^{-2}$ $\rm
s^{-1}$ with a photon index of $3.1 \pm 0.2$ in the period
2011-01-14 to 2011-01-19 UTC (Abdo et al. 2011b). Some other
analyses of LAT data were also done. A flare-like profile with a
flux (300 MeV -- 100 GeV) of around $4 \times 10^{-8}$ $\rm ph$ $\rm
cm^{-2}$ $\rm s^{-1}$ and a photon index of about 1.7 in the period
2010-11-18 00:00:00 (UT) to 2010-11-21 00:04:42 (UT) (Tam et al.
2010), and a flux (200 MeV -- 100 GeV) of $4.8 \pm 0.9 \times
10^{-7}$ $\rm ph$ $\rm cm^{-2}$ $\rm s^{-1}$ with a photon index of
$3.6 \pm 0.4$ in the period 2011-01-14 to 2011-01-16 UTC (Kong et
al. 2011) were reported.

\begin{figure}
\resizebox{\hsize}{!}{\includegraphics{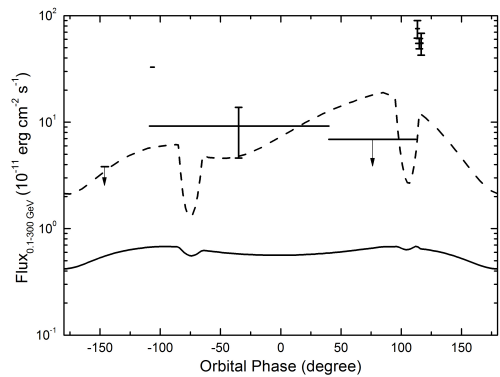}} \caption{The
calculated GeV light curves and their comparison with observations.
The emission in solid line is from synchrotron radiation and the
emission in dashed line is from the EIC process.}
\end{figure}

As shown in Figure 2, the emission between 100 MeV and 300 GeV range
is produced by the synchrotron and the EIC processes together. In
Figure 6, we present our calculated light curve between 100 MeV and
300 GeV using the parameters which are the same as those in the
dotted lines of Figure 1. The emission in solid line is from
synchrotron radiation and the emission in dashed line is from the
EIC process. The observation data are taken from The Astronomer's
Telegrams (ATel; Tam et al. 2010; Abdo et al. 2010a, 2010b, 2011b;
Kong et al. 2011), and the flux derived from Tam et al.'s (2010) and
Kong et al.'s (2011) reports have been extrapolated to the 100 MeV
and 300 GeV energy range using photon indices of 1.7 and 3.6
respectively. We can see that the calculated GeV flux are dominated
by the EIC process, but the EIC flux is still lower than the
observations. This may be due to two reasons: (1) In the ATel
reports, a single power-law has been assumed for the GeV photon
spectrum, and we use it to estimate the energy flux in the figure.
But in our calculation the spectrum is not a single power-law. 
It may overestimate the GeV flux by using a single power-law
spectrum in the figure. Some recent analyses actually show that the
spectrum between 100 MeV and 300 GeV is not a single power-law (Adbo
et al. 2011; Tam et al. 2011). (2) Some observations with high flux
may be flares, which cannot be produced in the steady model. In
other words the flares should be emitting from different region of
the shock, which is also suggested by Tam et al. (2011)
comparing the 100 MeV to 300 GeV spectra with the $>$ 300 GeV
gamma-ray spectra. We suggest that the flares may come from the
relativistic Doppler-boosting effect (Dubus, Cerutti \& Henri 2010).
Bogovalov et al. (2008) presented a hydrodynamic simulation of the
interaction between relativistic and nonrelativistic wind in PSR
B1259-63/SS 2883 system. They found that the electrons in the
shocked pulsar wind can be accelerated from bulk Lorentz factors
$\sim 1$ around the termination shock to nearly 100 far away. The
spectrum of synchrotron emission in our calculations peaks at about
0.1 GeV. With such a large bulk Lorentz factor of $\sim$ 100, the
peak will be Doppler-boosted to about 10 GeV, which could be
detected by $Fermi$ satellite. However, emission from the electrons
with a bulk Lorentz factor of 100 should be strongly beamed. When
the line-of-sight is in the beaming direction, we can receive the
GeV flux, otherwise, the GeV photons disappear. In this case, a
flare profile may be found. The much higher intensity of the flares
is mainly due to the beaming effect because instead of radiating
isotropically the GeV photons from the large bulk Lorentz factor
region is highly beamed at a much smaller solid angle. If the
geometry of the shock has a simple bow shock structure at the tail,
we expect that two flares will occur when the line of sight passes
through the two edges of the cone of the bow shock and each flare
can last about an orbital phase $2/\Gamma \sim
12^{\circ}(\Gamma/10)^{-1}$, where $\Gamma$ is the bulk Lorentz
factor at the tail of the shock. If the photon numbers are
sufficiently high during the flares we should be able to find the
synchrotron cut-off energy ($E_{\rm syn}\sim 100$ MeV cf. Figure 2)
boosting to $\Gamma E_{\rm syn} \sim 1$ GeV $(\Gamma/10)$ in these
flares and the general spectrum in $Fermi$ range should satisfy a
simple power with exponential cut-off.

Recently, Tam et al. (2011) found the flux in the flaring
period in Fermi observations is enhanced by a factor of 5-10, which
suggests a Doppler factor of around 1.5-2. With a low energy break
around 10 keV observed by $Suzaku$ (Uchiyama et al. 2009), the
enhancement has a factor of only about 3 in the X-ray band, which is
smaller than that in the GeV band and is less obvious. In addition,
the flaring period is near the second peak in the X-ray light curve.
This peak may be produced by the effect of Doppler boosting and disk
together. For the TeV flux, because more IC scattering occur in the
KN regime for $\Gamma > 1$, the TeV flux should be suppressed. In
addition, the stellar photon density in the bow shock tail is
smaller than that in the shock apex, so the IC emission should be
lower accordingly. The increase of TeV flux due to the Doppler
boosting and the decrease due to the KN effect and smaller photon
density will compensate each other, and the Doppler boosting may not
modulate the TeV light curve significantly. Furthermore, because the
emission from the shock tail is temporally and spectrally different
with that from the shock apex, the magnetic fields and the electron
distributions may not be the same in these two ranges.

\section{Conclusion and Discussion}

The PSR B1259-63/SS 2883 system is an attractive binary system, and
is also an unique astrophysical laboratory for probing the time
dependent interaction between the pulsar wind and stellar wind, and
studying the physics of the pulsar wind in a different range
compared with an isolated pulsar. In this paper, we have modeled the
X-ray and TeV observations of the source using the wind interaction
model under the hypothesis that the X-ray emission is mainly from
the synchrotron process and the TeV photons are mainly produced by
the EIC effect where the relativistic electrons in the shock
up-scatter the photons from the Be star. The effect of the disk
exhibits an emission and an absorption components in the X-ray and
TeV bands respectively. More importantly, we assume that the
microphysics parameters are not constant and take the magnetic
parameter $\sigma$ varying as a power-law form along with the
distance between the termination shock and the pulsar. This is a
reasonable assumption because $\sigma$ has a very large value at the
light cylinder and reaches a very small value of about 0.003 at a
distance of about 0.1 pc from the pulsar (Kennel \& Coroniti 1984a,
1984b). The variation of microphysics parameters has also been
suggested in some other astrophysical phenomena like Gamma-ray
bursts (Kong et al. 2010). We also  try to explain the GeV
light curve which is observed by LAT very recently, and suggest that
the actual photon spectra may not be power-law distributed as
reported in the ATel and some flare-like profile may come from the
relativistic Doppler-boosting effect.

In a recent paper, van Soelen \& Meintjes (2011) investigated
the effect of the infrare emission from the Be star's disk on the
inverse Compton process. The free-free and free-bound emission
occurred in the disk is thought to produce an infrared excess in the
Be star's photon spectrum. The scattering of this infrared excess
will influence the gamma-ray production. As discussed by van Soelen
\& Meintjes (2011), for the electrons with Lorentz factors
$\gamma_{\rm e} > 10^5$, the scattering of photons around the peak
in the stellar spectrum will be in the KN limit, while the
scattering with photons in the infrared excess will be in the
Thompson limit and increase the emission at GeV band. But for the
broad electron distribution with $\gamma_{\rm e,min} \sim 10^4$, the
scattering with photons around the peak of the stellar spectrum will
occur in the Thompson limit, and most of the GeV photons are produced
by the Thompson scattering of the whole stellar spectrum. So the
effect of the infrared excess on the spectrum could be negligible.
In our calculations above, the electron distributions are very broad
with ${\rm min} (\gamma_{\rm e,min},\gamma_{\rm e,c})$ between a few
$\times 10^3$ to a few $\times 10^4$. The resulting influence on the
GeV flux by the infrared excess is insignificant, so we do not
consider this effect in our calculations.

There are also some other binary systems in our Galaxy similar to
PSR B1259-63/SS 2883 being found for emitting the X-ray and VHE
emission, such as LS 5039 (Aharonian et al. 2006), LS
I+61\textordmasculine 303 (Albert et al. 2006) and possibly HESS
J0632+057 (Hinton et al. 2009). Our model can possibly be used in
explaining the broadband observational light curves in these
systems. However, the orbital properties are different in these
systems. In PSR B1259-63/SS 2883, the separation between the pulsar
and the massive star is 0.7-10 AU. But the separation is only about
0.1-0.2 AU in LS 5039 and 0.1-0.7 AU in LS I+61\textordmasculine 303
(Dubus 2006). The magnetization parameter $\sigma$ should be larger
accordingly. The detailed study of the radiation from the LS 5039
and the LS I+61\textordmasculine 303 systems could help us to
understand the physical condition of the pulsar wind in the position
where is much closer to pulsar compared with the PSR B1259-63/SS
2883 system.

\section*{Acknowledgments}

We would like to thank the anonymous referee for stimulating
suggestions that lead to an overall improvement of this study. We
also would like to thank Y. Chen, C. T. Hui, R. H. H. Huang, A. K.
H. Kong, P. H. T. Tam, J. Takata and X. Y. Wang for helpful
suggestions and discussion. This research was supported by a 2011
GRF grant of Hong Kong Government entitled "Gamma-ray Pulsars". YWY
is also supported by the National Natural Science Foundation of
China (Grant No. 11047121). YFH is supported by the National Natural
Science Foundation of China (Grant No. 10625313 and 11033002), the
National Basic Research Program of China (973 Program, Grant No.
2009CB824800).

\label{lastpage}

\end{document}